\documentclass[11pt]{article}
\usepackage{epsfig}
\usepackage{amsmath}
\usepackage{amssymb}
\usepackage{amsfonts}
\usepackage{enumerate}
\usepackage{cite}
\newcommand{\Hrule}{\vskip-10mm\begin{flushleft}\rule{\linewidth}{0.1mm}\end{flu
shleft}\vskip-1.5mm}
\title{Effect of Impurities in the Channel of a Spin Field Effect Transistor 
(SPINFET)}
\author{M. Cahay$^{\dag}$ and S. Bandyopadhyay$^{\dag \dag}$ \\
\\Department of Electrical and Computer Engineering and Computer 
Science\\
University of Cincinnati, 
Cincinnati, Ohio 45221--0030\\
Email: marc.cahay@uc.edu \\
\\$^{\dag \dag}$Department of Electrical Engineering\\
Virginia Commonwealth University, 
Richmond, Virginia 23284\\
Email: sbandy@vcu.edu}

\begin{document}
\maketitle
\twocolumn

\begin{abstract}
We show that the conductance of Spin Field Effect Transistors 
(SPINFET)
[Datta and Das, Appl. Phys. Lett., \underline{56}, 665 (1990)] is affected by 
 a single (non-magnetic) impurity in the transistor's channel.  The extreme 
sensitivity 
of the amplitude and phase of the transistor's conductance oscillations to the 
location of a single 
impurity in the channel is reminiscent of the phenomenon of universal 
conductance fluctuations 
in mesoscopic samples and is extremely problematic as far as device 
implementation is concerned. 
\end{abstract}

\section{{Introduction}}
\begin{figure}
\centering
\includegraphics[width=2.5in]{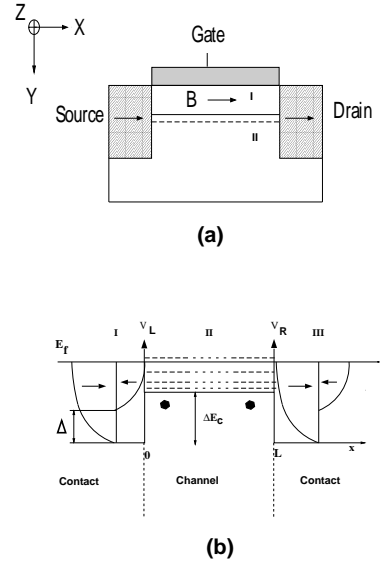}
\caption{Top: schematic of the electron spin interferometer
from ref. \cite{datta}.  The horizontal dashed line represents the quasi
one-dimensional electron gas formed at the semiconductor interface between
materials I and II.  The magnetization of the ferromagnetic contacts is assumed
to be along the +x-direction which results in a magnetic field along the
x-direction.  Bottom: Energy band diagram across the electron spin
interferometer.  We use a Stoner-Wohlfarth model for the ferromagnetic contacts.
$\Delta$ is the exchange splitting energy in the contacts.
$\Delta E_c$ is the height of the potential barrier between the energy band
bottoms of the semiconductor and the ferromagnetic electrodes. $\Delta E_c$ 
takes into
account the effects of the quantum confinement in the y- and z-directions.  Also 
shown
as dashed lines are the resonant energy states above $\Delta E_c$. Peaks in the
conductance of the electron spin interferometer are expected when the Fermi
level in the contacts lines up with the resonant states. The barriers at the
ferromagnet/semiconductor interface are modeled as simple one-dimensional
delta-potentials.}
\label{structure}
\end{figure}

In a seminal paper published in 1990, Datta and Das \cite{datta} proposed a
gate controlled electron spin interferometer which is an analog of the standard
electro-optic light modulator. Their device (later dubbed "Spin Field Effect 
Transistor" or SPINFET) consists of a  one-dimensional
semiconductor channel with ferromagnetic source and drain contacts (Fig. 
\ref{structure}(a)).
Electrons are
injected  into the channel from the ferromagnetic source with a definite spin 
orientation,
which is
then controllably precessed in the channel with a gate-controlled Rashba
interaction  \cite{rashba},
and finally  sensed
at the drain. At the drain end, the electron's transmission probability depends
on the relative alignment of its spin with  the drain's (fixed) magnetization.
By controlling
the angle of spin precession in the channel with a gate voltage,
one can modulate the relative spin alignment at the drain end, and hence control
the source-to-drain  current (or conductance). This is the principle of the 
SPINFET.

There have been some studies of spin transport in such a device. Mireles and
Kirczenow \cite{mireles1,mireles2} carried out a study of {\it ballistic} spin 
transport, 
but they overlooked two crucial features that are always present in a real
device structure. First, there is an axial magnetic field along the channel
caused by the ferromagnetic source and drain contacts which are magnetized in 
the same direction. This field can be quite strong ($\sim$ 1 Tesla) 
\cite{wrobel} and dramatically alters the
dispersion relations of the subbands in the channel, causes spin mixing, and has 
a serious effect on
spin transport.  Second, there will always be a few impurities in the channel
(even if they are remote impurities) associated with channel doping or 
unintentional defects. We show
that even a single (non-magnetic) impurity can cause spin relaxation
in the presence of the axial magnetic field.

\section{Theoretical Approach}
We first consider the quasi one-dimensional semiconductor channel of a SPINFET 
in the absence of any impurities.  The channel is along the
x-axis (Fig. \ref{structure}(a)) and the gate electric field is
applied along the y-direction to induce a Rashba spin-orbit coupling in the
channel. This
system  is described by the single particle effective-mass Hamiltonian
\cite{moroz}
\begin{eqnarray}
{\cal H} = {{1}\over{2 m^*}} \left ( {\vec p} + e{\vec A} \right )^2
+ V_I (x)
+ V_1 (y) + V_2 (z) \nonumber \\
-  (g^{*}/2) \mu_B {\vec B} \cdot {\vec \sigma} 
+  \frac{{\alpha}_R}{ \hbar} \hat{y}\cdot \left [ {\vec \sigma} \times ( {\vec
p} + e {\vec A} )
\right ]
\label{hamiltonian}
\end{eqnarray}
where $\hat{y}$ is the unit vector along the y-direction in
Fig. \ref{structure}(a) and ${\vec A}$ is the vector potential due to the axial 
magnetic
field ${\vec B}$ along the
channel (x-direction) caused by the ferromagnetic contacts. In (1), $
\mu_B $
is the Bohr
magneton ( $ e \hbar / 2 m_0$) and $g^* $ is the effective Land\'e g-factor of
the
electron in the channel.
The quantity
$\alpha_R$ is the Rashba spin-orbit coupling strength which  can be varied with
the gate potential.
The confining potentials along the y- and z-directions are denoted by $V_1 (y)$
and
$V_2 (z)$, with the latter being parabolic in space.

In (1), $V_I (x)$ represents an interfacial potential barrier
between the ferromagnetic contacts and the semiconducting channel. If the
contact neighborhood consists of heavily doped semiconductor material in close
proximity to a metallic ferromagnet,
the Schottky
barriers at the interface will be very narrow and electrons
from the contacts can tunnel fairly easily into semiconducting channel resulting
in a nearly-ohmic contact.
We model these ultra-narrow Schottky barriers as delta-barriers given
by:
\begin{equation}
V_I (x) = V_L \delta (x) + V_R \delta (x-L)
\end{equation}
where $ V_L $ and $V_R $ are assumed equal ($V_L$ = $V_R$ = $V_0$). In practice, 
the strength of the barrier depends on the ferromagnetic materials and also on 
the 
doping level in the channel.  These barriers have a beneficial effect; they can 
facilitate 
coherent spin injection across a metallic ferromagnet and a semiconducting 
paramagnet interface
\cite{eirashba1,heersche} 
which is crucial for the SPINFET.

In (1), we have neglected a few effects for the sake of
simplicity. We have neglected the normal Elliott-Yafet interaction 
\cite{elliott} because it is weak in quasi one-dimensional structures (where 
elastic scattering
is strongly suppressed). We have also neglected  the Dresselhaus
interaction \cite{dresselhaus} since it does not relax spin when the initial
spin polarization is along the axis of the wire \cite{bournel,pramanik}. The 
Dresselhaus interaction can however be easily included in the Hamiltonian  and 
is left for future work.

The choice of the Landau gauge ${\vec A}$ = (0, -Bz, 0)  allows us to decouple
the y-component of the Hamiltonian in (1) from the x-z component.
Furthermore, in the absence of any impurity scattering potential ($V_{imp} = 
0$), the Hamiltonian in the semiconducting channel is translationally invariant 
in the x-direction and the wavevector $k_x$ is a good quantum number. The 
eigenstates
of the system can then be determined using plane waves traveling in the 
x-direction \cite{losalamos}.  
The two-dimensional Hamiltonian in the plane of the channel (x-z plane) is 
therefore given by
\begin{eqnarray}
H_{xz} & = & {{p_z^2}\over{2 m^*}} + \Delta E_c + V_I (x) + {{1}\over{2}}m^* 
\left (
\omega_0^2 +
\omega_c^2 \right ) z^2 + {{\hbar^2 k_x^2}\over{2 m^*}} \nonumber \\
& & +{{\hbar^2 k_R
k_x}\over{m^*}} \sigma_z - ( g^* /2) \mu_B B {\sigma}_x  - {{\hbar k_R
p_z}\over{m^*}} \sigma_x
\label{Hamiltonian}
\end{eqnarray}
where $\omega_0$ is the curvature of the confining potential in the z-direction,
 $k_R = m^*
\alpha_R/\hbar^2$, and $\Delta E_c$ is the potential barrier between the
ferromagnet and
semiconductor. We assume that $\Delta E_c$ includes the effects of the quantum
confinement in the y-direction (Fig.1(a)).

To model localized non-magnetic impurities (i.e., which do not by themselves 
flip the spin) we use a standard model of
delta-scatterers
\begin{equation}
V_{imp} = {\Gamma}_{imp} \delta ( x - x_i)
\end{equation}
for an impurity located at a distance $x_i$ from the
left ferromagnet/semiconductor interface with scattering strength
${\Gamma}_{imp}$ (assumed to be spin independent). In our numerical
examples, we consider the case of both attractive (${\Gamma}_{imp}$ negative) 
and
repulsive (${\Gamma}_{imp}$ positive) impurities. While (1) and (3) represent a
ballistic channel with no scattering, addition of the scattering potential in
(4) to (1) or (3) will result in a Hamiltonian describing a weakly
disordered channel in which impurity scattering takes place. The eigenstates of
this (spin-dependent) Hamiltonian can then be found using a transfer matrix
technique to extract the electron wavefunction. From this wavefunction, we can
calculate the (spin-dependent) transmission probability through the channel and
ultimately the (spin-dependent) channel conductance. The details of the
calculations have been presented elsewhere \cite{losalamos}.  The linear 
response conductance of 
the spin interferometer (for injection from either the +x or -x polarized bands 
in 
the left contact) is found from the Landauer formula.

We model the ferromagnetic contacts by
the Stoner-Wohlfarth model (Fig.1.(b)). The
+x-polarized spin (majority carrier) and -x-polarized spin (minority carrier)
band bottoms are offset by an exchange splitting energy $\Delta$.
The solutions of the Schr\"odinger
equation for injection from minority and majority spins from the left magnetic 
contact
then be written throughout the entire device leading to the
corresponding transmission probabilities and the conductance of the 
interferometer based on the Landauer formula (see \cite{losalamos}). The 
strength of the barrier at the ferromagnet/semiconductor
interface is characterized by the following parameter
\begin{equation}
Z = \frac{ 2 {m_f}^* V_0 }{ {\hbar}^2}.
\end{equation}
Typical values of $Z$ vary in the range of 0 to 2 \cite{schapers2}.
Using ${m_f}^* = m_0$ and $k_F $ = 1.05x$10^8 $ cm$^{-1}$, we get a barrier
strength $V_0$ = 16 eV$\AA$ for Z = 2.

\section{Numerical Examples}
We consider a spin interferometer consisting of
a quasi one-dimensional InAs channel between two ferromagnetic contacts.  The
electrostatic potential in the
z-direction is assumed to be harmonic (with $\hbar \omega_0$ = 10 meV in (3).
A Zeeman splitting energy of 0.34 meV is used in the semiconductor channel
assuming a magnetic field B = 1 Tesla along the channel.
This corresponds to a  $g^*$ factor of 3 and an electron effective mass $m^* =
0.036 m_o$ which is typical of InAs-based channels \cite{datta}. The Fermi level
$E_f$ and the exchange splitting energy $\Delta $ in the ferromagnetic contacts
are set equal to 4.2 and 3.46 eV, respectively  \cite{mireles}.

The Rashba spin-orbit coupling strength ${\alpha}_R$ is
typically derived from low-temperature magnetoresistance measurements
(Shubnikov-de Haas oscillations) in 2DEG created at the interface of
semiconductor heterostructures \cite{nitta}. To date, the largest reported
experimental values of the Rashba spin-orbit coupling strength ${\alpha}_R$ has
been found in
InAs-based semiconductor heterojunctions. For a normal HEMT
$In_{0.75} Al_{0.25} As/In_{0.75} Ga_{0.25} As $ heterojunction, Sato et al.
have reported variation
of ${\alpha}_R$ from 30- to 15 $\times 10^{-12}$ eV-m when the external gate
voltage is swept from 0 to -6 V (the total electron
concentration in the 2DEG is found to be reduced from 5- to 4.5$\times 10^{11}
/cm^2$ over the same range of bias).

Tuning the gate voltage varies both  the potential
energy barrier $\Delta E_c$ and the Rashba spin-orbit coupling strength
$\alpha_R$. Both of these variations lead to distinct types of conductance
oscillations. The variation of $\Delta E_c$ causes the Fermi-level in the
channel
to sweep through the resonant energies in the channel, causing the conductance
to oscillate. These are known as Ramsauer oscillations and have been examined by
us in detail in \cite{losalamos}. The variation of $\alpha_R$, on the other
hand, causes spin precession in the channel leading to the type of
conductance oscillation which is the basis of the spin interferometer, as
originally visualized by Datta and Das \cite{datta}. In \cite{losalamos} we 
found that 
the Ramsauer oscillations are much stronger and can mask the oscillations due to 
spin precession, 
unless the structure is designed with particular care to eliminate (or reduce) 
the Ramsauer
oscillations.

In the numerical examples below, we assumed that the Rashba spin-orbit
coupling strength ${\alpha}_R$ is constant (i.e., independent of the gate 
potential)
and equal to the maximum reported value of 30 $\times 10^{-12}$ eV-m (best case 
scenario).
We also used a value of Z = 0.25 corresponding to a value of $ V_L $
and $V_R$ in (2) equal to 2 eV-$\AA$. The scattering strength of the single
impurity in the channel was set equal to $\Gamma$ = $\pm$ 0.25 eV-$\AA$ (the
plus and minus sign corresponding to repulsive and attractive impurity, 
respectively).

First, we consider the case of a single
repulsive scatterer (impurity) whose location is varied from the left to the 
rigth side of the channel
in steps of 10 $\AA$.  Figure 2 shows that variation of the conductance with the 
impurity location for three 
different values of $\Delta E_c $ in the channel, i.e, gate potential on the 
SPINFET. 
It is seen that the condutance of the interferometer is a strong function of the 
impurity location. 
Figure 2 shows that the condutance modulation
(difference between maximum and minimum conductance values) of the 
interferometer is about
0.9, 0.21, and 0.2 $ e^2 /h$, for $\Delta E_c$ equal to 4.191, 4.188, and 4.185 
eV, respectively.
The same calculation were repeated for the case of an attractive impurity with 
the
same magnitude of the scattering strength and the conductance modulation as a 
function of
the impurity location are shown in Fig.3.  In this case, the condutance 
modulation
(difference between maximum and minimum conductance values) of the 
interferometer is about
0.76, 0.16, and 0.17$ e^2 /h$, for $\Delta E_c$ equal to 4.191, 4.188, and 4.185 
eV, respectively.

\vskip .2in
\section{Conclusions:}
In this paper, we have shown
how the conductance of gate controlled spin interferometers proposed in
\cite{datta} is strongly dependent on the location
of a single impurity in the semiconducting channel (Figs. 2 and 3).   The 
extreme sensitivity of the amplitude
and phase of conductance oscillations of a SPINFET to impurity location is 
reminiscent of the phenomenon of
universal conductance fluctuations of mesoscopic samples \cite{stone}. This will 
hinder practical
applications of electron spin interferometers since it will lead to such 
problems as large threshold variability, random device characteristics, and 
general irreproducibility.

The work of S. B. was supported by the National Science Foundation.

\begin{figure}
\centering
\includegraphics[width=2.5in]{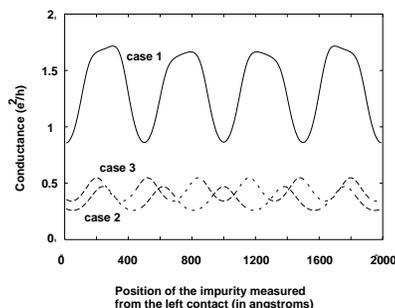}
\caption{ Conductance modulation of a SPINFET as a function of the position
of a single repulsive impurity from the left ferromagnet/semiconductor 
interface.
The channel is 2000 $\AA$ long. The strength of the delta-scatterer
${\Gamma}_{imp}$ is set equal to 0.25 eV$\AA$.
The three curves correspond to different values of the
potential barrier height $\Delta E_c$ which is controlled by the gate voltage.
Case 1, 2, and 3 corresponds to a value of $\Delta E_c$ equal to
4.188, 4.185, and 4.191 eV, respectively. The calculations are for
absolute zero temperature.}
\end{figure}

\begin{figure}
\centering
\includegraphics[width=2.5in]{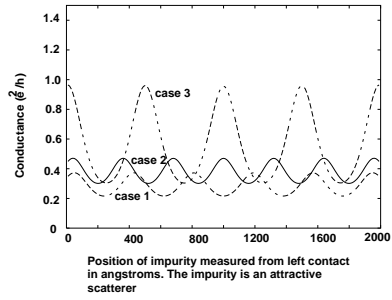}
\caption{Same as Fig.2 for an attractive scatterer
with scattering strength ${\Gamma}_{imp}$ = -0.25 eV$\AA$.
Case 1, 2, and 3 corresponds to a value of $\Delta E_c$ equal to
4.191, 4.185, and 4.188 eV, respectively. The calculations are for
absolute zero temperature.}
\end{figure}


\end{document}